\documentclass[a4paper,11pt]{article}

\usepackage[utf8]{inputenc}
\usepackage[UKenglish]{babel}
\usepackage[p,osf]{cochineal}
\usepackage[varqu,varl,var0]{inconsolata}
\usepackage[scale=.95,type1]{cabin}
\usepackage[cochineal,vvarbb]{newtxmath}
\usepackage[scr=boondoxo]{mathalfa}
\usepackage[mathcal]{euscript}
\usepackage[dvipsnames]{xcolor}

\usepackage[top=2.5cm, bottom=3cm, left=2.6cm, right=2.6cm]{geometry}
\usepackage{setspace}
\makeatletter
\g@addto@macro\bfseries{\boldmath}
\makeatother
\usepackage{fancyhdr}
\usepackage{titlesec}
\titleformat{\subsubsection}{\bfseries}{\thesubsubsection}{1em}{}                    
\titlespacing*{\subsubsection}{0pt}{0.5\baselineskip}{0.2\baselineskip}

\usepackage{amsmath,amsthm}
\usepackage{amsfonts}
\usepackage{bm}
\usepackage{slashed}
\usepackage{mathtools}
\usepackage{amsthm}
\usepackage{amsbsy}
\numberwithin{equation}{section}
\allowdisplaybreaks
\usepackage{mdframed}
\usepackage{pifont}
\usepackage{empheq}

\usepackage{pict2e}
\makeatletter
	\newcommand{\loplus}{\mathbin{\mathpalette\dog@lsemi{+}}}
	\newcommand{\dog@lsemi}[2]{\dog@semi{#1}{#2}{270,90}}
	\newcommand{\dog@semi}[3]{%
	\begingroup
		\sbox\z@{$\m@th#1#2$}%
		\setlength{\unitlength}{\dimexpr\ht\z@+\dp\z@\relax}%
		\makebox[\wd\z@]{\raisebox{-\dp\z@}{%
    		\begin{picture}(1,1)
    			\linethickness{\variable@rule{#1}}
    			\roundcap
    			\put(0.5,0.5){\makebox(0,0){\raisebox{\dp\z@}{$\m@th#1#2$}}}
    			\put(0.5,0.5){\arc[#3]{0.5}}
    		\end{picture}%
  		}}%
	\endgroup
	}
	\newcommand{\variable@rule}[1]{%
 	\fontdimen8  
  	\ifx#1\displaystyle\textfont3\else
    \ifx#1\textstyle\textfont3\else
    \ifx#1\scriptstyle\scriptfont3\else
    \scriptscriptfont3\relax
  	\fi\fi\fi
	}
\makeatother
	
\let\oldsqrt\sqrt
\def\sqrt{\mathpalette\DHLhksqrt}
\def\DHLhksqrt#1#2{%
\setbox0=\hbox{$#1\oldsqrt{#2\,}$}\dimen0=\ht0
\advance\dimen0-0.2\ht0
\setbox2=\hbox{\vrule height\ht0 depth -\dimen0}%
{\box0\lower0.4pt\box2}}

\newmdenv[
	innerleftmargin=10pt,
  	innerrightmargin=10pt,
  	innertopmargin=10pt,
  	innerbottommargin=10pt,
  	skipbelow=5pt,
  	leftmargin=0pt,
  	rightmargin=0pt,
  	linecolor=black,
  	linewidth=0.5pt
]{defbox}

\usepackage{enumitem}
\setlist[itemize,1]{label=$\rhd$}
\setlist[itemize,2]{label=$\bullet$}
\usepackage{array}
\usepackage{multirow}

\usepackage{tabularx}
\newcolumntype{C}[1]{>{\centering\arraybackslash}p{#1}}
\usepackage{colortbl}
\usepackage{boldline}

\usepackage[textfont=it]{caption} 

\usepackage{hyperref}
\hypersetup{
    colorlinks,%
    citecolor=blue,%
    filecolor=blue,%
    linkcolor=blue,%
   	urlcolor=blue,
   	linktoc=page
}

\usepackage{cite}

\usepackage[toc,page]{appendix}

\usepackage{graphicx}
\usepackage{subcaption}
\usepackage{float}
\usepackage{tikz}
\usepackage{tikz-layers}
\usepackage{tikz-3dplot}
\usetikzlibrary{calc}
\usetikzlibrary{decorations.pathreplacing}
\usetikzlibrary{decorations.pathmorphing}
\usetikzlibrary{decorations.markings}
\usetikzlibrary{arrows.meta}
\usetikzlibrary{positioning}
\usetikzlibrary{math}
\usetikzlibrary{cd}
\tikzset{every picture/.style={font issue=\footnotesize},
         font issue/.style={execute at begin picture={#1\selectfont}}
        }

\newcommand{\D}{\mathop{}\!\text{d}}

\newcommand{\bfupsilon}{{\boldsymbol{\upupsilon}}}

\newcommand{\bfg}{{\text{\bf g}}}

\newcommand{\vect}[1]{{\text{\bf #1}}}

\usepackage{comment}

\begin{document}


\begin{titlepage}

\begin{flushright}
{\small \hfill \today}
\end{flushright}
\vspace{0.5cm}

\begin{center}
{\textbf{\textsc{
	{\LARGE Kinetic Theory of Carroll Hydrodynamics}
}}}

\end{center}

\vspace{7mm}
\begin{center} 
Victor Chabirand$^{a,}$\footnote{E-mail: \href{mailto:victor.chabirand@polytechnique.edu}{victor.chabirand@polytechnique.edu}}, Adrien Fiorucci$^{a,b,}$\footnote{E-mail: \href{mailto:adrien.fiorucci@polytechnique.edu}{adrien.fiorucci@polytechnique.edu}}, P. Marios Petropoulos$^{a,}$\footnote{E-mail: \href{mailto:marios.petropoulos@polytechnique.edu}{marios.petropoulos@polytechnique.edu}} and Matthieu Vilatte$^{c,}$\footnote{E-mail: \href{mailto:matthieu.vilatte@umons.ac.be}{matthieu.vilatte@umons.ac.be}} 

\normalsize
\bigskip\bigskip
$^{a}$ \textit{Centre de Physique Th\'eorique, \'Ecole polytechnique, \\
Centre National de la Recherche Scientifique -- Unit\'e Mixte de Recherche 7644, \\
Institut Polytechnique de Paris, 91120 Palaiseau Cedex, France}
        
$^b$ \textit{Université Libre de Bruxelles and International Solvay Institutes \\
CP 231, Boulevard du Triomphe, 1050 Bruxelles, Belgium}

$^c$ \textit{Service de Physique de l’Univers, Champs et Gravitation, Université de Mons \\
20 Place du Parc, 7000 Mons, Belgium}

\vspace{30pt}

\begin{center}
	\begin{minipage}{0.85\textwidth}
		\begin{abstract}\noindent
			We develop the foundations of Carrollian statistical mechanics by considering a system of interacting instantonic space-filling branes on a flat background, thereby providing the closest Carrollian analogue to the Galilean gas of interacting particles that underpins Boltzmann’s collision theory. By adapting Boltzmann’s statistical approach within this framework, we provide a first-principles microscopic derivation of the so-called Carrollian fluid equations, which were previously obtained as the vanishing-speed-of-light limit of relativistic conservation laws. We then use this analysis as a basis for formulating the first elements of Carrollian thermodynamics.
		\end{abstract}
	\end{minipage}
	
	\bigskip
	{\small Pre-print number: CPHT-RR015.052026}
\end{center}

\end{center}

\end{titlepage}



\setcounter{tocdepth}{2}
\begingroup
\setlength{\parskip}{1pt} 
\tableofcontents
\endgroup

\vspace{15pt}
\noindent\hrulefill
\setcounter{footnote}{0} 

\setstretch{1.15}


\section{Introduction}

Over the centuries, three types of relativity have been introduced. Galilean relativity describes our everyday experience. Einstein relativity has proven useful for accurately accounting for phenomena occurring on astrophysical and cosmological scales, as well as high-energy interactions such as collider physics. Devised in 1965 by Lévy-Leblond \cite{Levy1965}, Carroll relativity was long regarded as a mathematical curiosity. While the Galilean regime is reached by formally taking the infinite-speed-of-light limit of Einstein relativity, Carroll physics can be understood as its vanishing-speed-of-light limit. It is now taken more seriously because of alleged applications in condensed-matter physics \cite{Bidussi:2021nmp,Bagchi:2022eui,Figueroa-OFarrill:2023vbj,Figueroa-OFarrill:2023qty,Kasikci:2023tvs,Ara:2024fbr,Biswas:2025dte} and cosmology \cite{deBoer:2021jej,Oling:2024vmq,Blair:2025nno,Argandona:2025jhg}, but above all because of its emergence at the conformal boundary of asymptotically flat spacetimes in a holographic setting directly inspired by the AdS/CFT correspondence \cite{Duval:2014lpa,Duval:2014uva,Bagchi:2016bcd,Penna:2017vms,Ciambelli:2018ojf,Ciambelli:2018wre,Ciambelli:2019lap,Ciambelli:2020eba,Campoleoni:2022wmf,Mittal:2022ywl,Donnay:2022aba,Donnay:2022wvx,Campoleoni:2023fug,Mason:2023mti,Alday:2024yyj,Lipstein:2025jfj,Fiorucci:2025twa,Hartong:2025jpp}. However, it is fair to say that many basic aspects of Carroll physics remain poorly understood and require careful examination in view of the aforesaid applications.

Hydrodynamics provides a paradigmatic example of this general phenomenon. Although the collapse of light cones inherent in the Carroll limit makes the very notion of a fluid elusive, the Carroll limit of relativistic fluid dynamics has been investigated. The latter has been instructive, as it has shed light on the mathematical structure of the fluid equations \cite{Athanasiou:2024lzr,Athanasiou:2024ykt,Petropoulos:2024jie}, the systematic exchange of time and space derivatives and momenta when comparing Galilean and Carrollian regimes \cite{Ciambelli:2018xat,Ciambelli:2018wre}, the fate of hydrodynamic-frame invariance \cite{Petkou:2022bmz}, and recurrent dynamical features such as the absence of matter-driven energy flux \cite{deBoer:2017ing,deBoer:2021jej}. Yet an intrinsic and physical understanding of these so-called \textit{Carroll fluid equations} has been missing for almost a decade. This nevertheless remains an important open question that has sustained the community's interest. It is indeed desirable to gain some understanding of the collective dynamics of degrees of freedom in the Carroll limit, chiefly because of its application to flat-space holography, where Carroll fluids have been shown to reproduce bulk Einstein dynamics from intrinsic boundary data \cite{Fiorucci:2025twa,Hartong:2025jpp}.  

In this paper, we address and solve this question by offering a first-principles derivation of the Carroll fluid equations from a statistical analysis. We take inspiration from Boltzmann's pioneering approach to the collective motion of a gas of particles arising from internal randomised molecular interactions \cite{Pitaevskii1981Physical,Huang1987,2002rbet.book.....C,Rezzolla}. The subtlety that arises when translating this to Carroll systems is that local excitations are immobile in this regime, because energy is a central element of the Carroll algebra. However, there are other interesting representations of this algebra that allow for instantaneous motion in space, and are better understood in terms of instantonic space-filling branes $t(\vect x)$. Recalling the duality between the two limits, they are the appropriate objects with which to build Carrollian mechanics, as counterparts of massive Galilei particles that follow trajectories $\vect x(t)$ over space. This observation was made in the seminal work \cite{Ciambelli:2018xat} but was never pursued thereafter; it nevertheless represents the required change of paradigm that one must concede in order to understand the microscopic degrees of freedom underlying what one might call \textit{Carroll hydrodynamics}.

Our analysis proceeds as follows. First, we define a system of interacting space-filling branes on the flat Carroll manifold and derive the related equations of motion, which can be elevated to the fundamental principles of Carroll mechanics. We thereby introduce the instrumental notion of spacelike evolution and the associated conserved quantities. Next, we establish the collision theory within this system of branes due to random internal excitations and show how the Carroll fluid equations arise upon averaging over these excitations. This leads us to a natural notion of temperature and entropy, on which we comment in the last part of the article. Our findings open the way to an intrinsic dynamical study of the Carroll limit with many virtues. First, our approach exploits to the full the peculiarities of Carroll physics without seeking to mimic the well-known Galilean and Einsteinian cases. Second, it avoids by construction the issues of regularity and well-definedness in Carroll thermodynamics that have been raised in the literature, see \textit{e.g.}\ \cite{deBoer:2023fnj} and sets the stage for a better understanding of flows on null hypersurfaces together with their thermal properties, which are of direct relevance to flat-space holography and black-hole physics.

\section{System of interacting instantonic space-filling branes}
\label{sec:mechanics}

We work on the $(d+1)$-dimensional flat Carrollian manifold $(\mathcal M,\bfg,\bfupsilon)$ \cite{Henneaux:1979vn,Duval:2014uoa,Bekaert:2015xua} covered by Cartesian coordinates $(t,\vect x)$, for which the degenerate metric and the field of observers read as
\begin{equation}
	\bfg = 0\times \D t^2 + \delta_{ij}\D x^i\D x^j,\quad \bfupsilon = \partial_t,\quad i,j = 1,\dots,d. \label{eq:flat}
\end{equation}
The comparison between Galilean and Carrollian dynamics reveals that these two limits of special relativity are related by a duality exchanging the roles of time and space. Building on this intuition, the natural dynamical object to consider on Carroll backgrounds appears to be a space-filling brane $t(\vect x)$ rather than a particle following a trajectory $\vect x(t)$ \cite{Ciambelli:2018xat}. We denote by $u_i(\vect x)\coloneqq D_i t(\vect x)$ the local gradient of the field $t(\vect x)$, where
\begin{equation}
	D_i \coloneqq \frac{\partial}{\partial x^i} + u_i \frac{\partial}{\partial t} + D_i u_j \frac{\partial}{\partial u_j} + \text{higher-order terms}
\end{equation}
is the total derivative with respect to $x^i$.\footnote{Since the ``evolution'' parameter is here replaced by the spatial coordinates $\vect x$, the total derivative $D_i$ with respect to space is the analogue of the total time derivative in classical mechanics, usually denoted $\dot F = \partial_t F + \dot x \partial_x F$. See \textit{e.g.}\ \cite{Olver1986ApplicationsOL} for a review.} The covector $\vect u = u_i\D x^i$ replaces the instantaneous velocity of Galilean mechanics and has dimensions of time over length. It constitutes the natural kinematic quantity that survives the Carroll limit \cite{Petkou:2022bmz}; we call it the \textit{local inverse velocity}. Importantly, iterated total derivatives $D_{i_1} D_{i_2}\dots D_{i_k} t$ are fully symmetric spacelike tensors by construction, since $t$ is assumed to be smooth.\footnote{One can indeed show that total derivatives commute while acting on smooth functions, \textit{i.e.}\ $D_i D_j f = D_j D_i f$, see again \cite{Olver1986ApplicationsOL}.}

We consider a collection of $N$ such branes $\{t_a(\vect x)\}_{a=1}^N$, whose dynamics is governed by the action
\begin{equation}
	S = \int \left( \frac{1}{2} \sum_a \sigma\delta^{ij}u_i^a(\vect x) u_j^a(\vect x) - V\big(\{t_a(\vect x)\}\big) \right) \D^d x. \label{eq:action carr}
\end{equation}
For each $a$, the kinetic term in this action can be obtained as the vanishing-speed-of-light limit of the Nambu--Goto\footnote{This action has also been referred to as the free Dirac--Born--Infeld action.} action for a relavistic $d$-brane \cite{Ciambelli:2018xat}. The potential $V$ depends on the ``temporal positions'' $t_a$ of the branes, but neither on the inverse velocities nor explicitly on the spatial coordinates, so as to mimic conservative Galilean systems. The constant $\sigma$, which describes the branes' isotropic resistance to spatial deformations,\footnote{This amounts to neglecting internal degrees of freedom on the branes. In this analysis, the latter are considered isotropic and homogeneous, and their local strain tensor is proportional to the identity, $\sigma_{ij}=\sigma \delta_{ij}$. The study of more general configurations with arbitrary strain tensors $\sigma_{ij}\neq \sigma \delta_{ij}$ can also be carried out.} is assumed to be universal, since we wish to consider a monospecies system. The equations of motion
\begin{equation}
D_i(\sigma D^i t_a) = -\frac{\partial V}{\partial t_a} \coloneqq P_a, \label{eq:eom}
\end{equation}
for each brane, provide the \textit{fundamental principle of Carrollian mechanics}: the spatial variation of the inverse velocity is driven by the total Carrollian power $P_a$ exerted on the system.

As in classical mechanics, Noether's first theorem asserts that continuous symmetries of the action \eqref{eq:action carr} lead to conserved quantities $Q$ along trajectories. Since our variable is now $\vect x$ rather than $t$, conservation should no longer be understood as $\dot Q=0$, but instead as $D_i Q^i = 0$ when the equations of motion \eqref{eq:eom} are satisfied. We are thus led to seek currents that are spatially conserved. Assuming that there is no external potential, \textit{i.e.}\ that $V$ only models mutual interactions and therefore satisfies $V = V\big(|t_a-t_b|\big)$, the action \eqref{eq:action carr} is invariant under time translation ($t'=t+a$, $\vect x'=\vect x$), which yields the on-shell conservation of the energy current
\begin{equation}
	\bar E^i = \sum_a E^i(\vect u_a),\quad E^i(\vect u) \coloneqq \sigma u^i. \label{eq:energy current}
\end{equation}
We decide to keep dubbing ``energy'' the conserved quantity associated with time-translation invariance: this also justifies that $P_a$, introduced in Eq.\ \eqref{eq:eom}, can be seen as a power as it quantifies the flux of energy upon space evolution. More comments on our choices of units and nomenclature can be found in Appendix \ref{app:carrvsgal}. Space-translation invariance ($t'=t$, $\vect x' = \vect x+\vect a$) implies that the stress tensor
\begin{equation}
	\bar T^i{}_j = \sum_a K^i{}_j(\vect u_a) + V \delta^i{}_j,\quad K_{ij}(\vect u) \coloneqq \sigma \big(u_i u_j - \tfrac12 \delta_{ij}\vect u^2\big), \label{eq:stress tensor}
\end{equation}
is conserved, $D_i \bar T^i{}_j=0$, when \eqref{eq:eom} holds. Furthermore, the isometries of the flat Carroll manifold \eqref{eq:flat} are fully accounted for by also considering spatial rotations ($t'=t$, $\vect x' = R\vect x$, $R\in SO(d)$) and Carroll boosts ($t'=t-\vect b\cdot\vect x$, $\vect x'=\vect x$) for some constant covector $\vect b$. These symmetries provide two additional Noether charges: $\bar M^i{}_{jk} = \bar T^i{}_{[j}x_{k]}$ and $\bar B^i{}_j = \sigma\delta^i{}_j(\sum_a t_a) - \bar E^i x_j$, which respectively obey $D_i \bar M^i{}_{jk}=0$ and $D_i \bar B^i{}_j=0$ on shell.

\section{Collision theory}
\label{sec:collision}

We now discuss brane interactions in order to initiate the Carrollian counterpart of Boltzmann's collision theory. In what follows, we assume that the branes under consideration are completely characterised by the pair $(t_a,\vect u_a)$ of functions of space, and that there are no internal degrees of freedom to be taken into account. The core of Boltzmann's analysis is the description of local binary collisions, which must, of course, be refined in the case of extended objects.

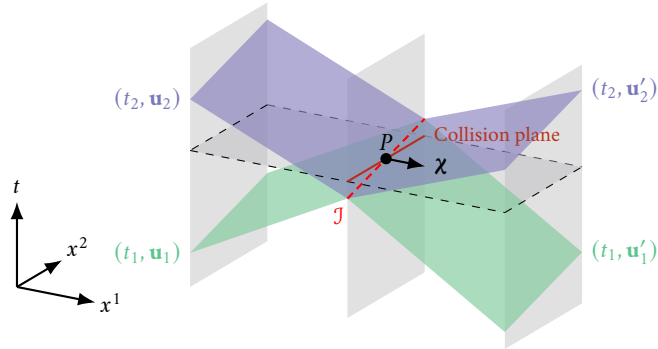
\begin{figure}[ht]
    \centering
    \tdplotsetmaincoords{70}{30} 
    \begin{tikzpicture}[tdplot_main_coords,scale=1.2]
        \fill[fill=Black!30!white,opacity=0.45] (0,0,1.2) -- (0,1.7,1.2) -- (4,1.7,1.2) -- (4,0,1.2);
        
        \draw[dashed, black] (0,1.7,1.2) -- (4,1.7,1.2) ;

        \fill[fill=Black!30!white,opacity=0.45] (0,0,-0.4) -- (0,1.7,-0.4) -- (0,1.7,2.4) -- (0,0,2.4);
        \fill[fill=Black!30!white,opacity=0.45] (2,0,-0.4) -- (2,1.7,-0.4) -- (2,1.7,2.4) -- (2,0,2.4);
        \fill[fill=Black!30!white,opacity=0.45] (4,0,-0.4) -- (4,1.7,-0.4) -- (4,1.7,2.4) -- (4,0,2.4);

        \fill[fill=Green!50!white,opacity=0.65] (0,0,0) node[left, Green] {$(t_1,\vect u_1)$} -- (0,1.7,0.4) -- (2,1.7,1.4) -- (2,0,1);
        \fill[fill=Green!50!white,opacity=0.65] (4,0,-0.2) -- (4,1.7,0.2) node[right, Green] {$(t_1,\vect u'_1)$} -- (2,1.7,1.4) -- (2,0,1);

        \fill[fill=Blue!50!white,opacity=0.65] (0,0,1.8) node[left, Blue] {$(t_2,\vect u_2)$} -- (0,1.7,2.2) -- (2,1.7,1.4) -- (2,0,1);
        \fill[fill=Blue!50!white,opacity=0.65] (4,0,1.7) -- (4,1.7,2.1) node[right, Blue] {$(t_2,\vect u'_2)$} -- (2,1.7,1.4) -- (2,0,1);

        \draw[dashed, black] (0,0,1.2) -- (4,0,1.2);

        \draw[densely dashed, red, thick] (2,0,1) node[anchor=north east,xshift=2pt]{$\mathcal I$} -- (2,1.7,1.4);
        \draw[thick, BrickRed] (2,0,1.2) -- (2,1.7,1.2) node[right] {\scriptsize Collision plane};
        
        \draw[-Latex, thick, black] (2,0.85,1.2) node[above] {$P$} -- (2.5,0.85,1.2) node[right] {$\bm{\upchi}$};
        
        \node[circle,inner sep=1.5pt,fill] at (2,0.85,1.2){};
        
        \draw[dashed, black] (0,1.7,1.2) -- (0,0,1.2) node[left] {};
        \draw[dashed, black] (4,0,1.2) -- (4,1.7,1.2) node[right] {};

        \draw[-Latex, thick] (-2.2,0,-0.8) -- (-2.2,0,0.2) node[above] {$t$};
        \draw[-Latex, thick] (-2.2,0,-0.8) -- (-1.2,0,-0.8) node[right,inner sep=1pt] {$x^1$};
        \draw[-Latex, thick] (-2.2,0,-0.8) -- (-2.2,1,-0.8) node[anchor=south west,inner sep=1pt] {$x^2$};
    \end{tikzpicture}
    \caption{Collision between two space-filling branes.
    }
    \label{fig: collision between two branes}
\end{figure}

Consider two branes $(t_1,\vect u_1)$ and $(t_2,\vect u_2)$ that interact over a region $\mathcal I\subset\mathbb R^d$ as a result of the system's internal agitation. We assume that large overlaps and recombinations are exceedingly unlikely and therefore constitute a null set in configuration space. This implies that $\mathcal I$ has codimension at least two, \textit{i.e.}\ it is at most $(d-1)$-dimensional, therefore excluding condensation and other large-scale effects. Moreover, the probabilistic set of contacts along lower-dimensional manifolds is also null. Therefore, restricting attention to the most probable configurations, we approximate the kinematics of interactions by assuming that such processes occur only along codimension-two manifolds, and that the resulting geometric perturbations dissipate sufficiently rapidly away from $\mathcal I$ to be treated as quasi-local. We refer to these binary quasi-local processes as \textit{brane collisions}. Fig.~\ref{fig: collision between two branes} depicts an example of such a collision.

By locality, there always exists a sufficiently small neighbourhood $\mathcal U_P$ of any point $P\in\mathcal I$ in which no other brane collision occurs. At the expense of reducing $\mathcal U_P$, we can assume that $\mathcal U_P\cap \mathcal I$ is a $(d-1)$-dimensional hyperplane. Its projection onto the constant-$t$ hyperplane through $P$ then defines a horizontal codimension-two hyperplane, with unit normal vector $\bm\upchi$,\footnote{The explicit expression of $\bm\upchi$ in terms of the inverse velocities $\vect u_1$ and $\vect u_2$ reads as $\bm\upchi = \pm(\vect u_2-\vect u_1)/|\vect u_2-\vect u_1|^2$.} which we call the \textit{collision hyperplane}. The orientation of $\bm\upchi$ is entirely arbitrary; once fixed, however, it provides a notion of ``initial'' and ``final'' states on the flow of $\bm\upchi$. This is how our interpretation of Carrollian ``evolution'' along spatial directions acquires its full meaning. Under these assumptions, the brane system behaves locally in a manner sufficiently close to that of a Boltzmann system for the same methods to be applied in establishing its statistical mechanics.

We begin by introducing the distribution function $f(\vect x,t,\vect u)$ such that the number of branes at the point $\vect x$, lying within an infinitesimal time interval $(t,t+\D t)$ and with inverse velocities ranging from $\vect u$ to $\vect u+\D\vect u$, is, by definition, given by $f(\vect x,t,\vect u)\D t \D^d u$. The local temporal density $n=n(\vect x,t)$ and the total number $N$ of branes are then obtained from $f$ as
\begin{equation}
	n = \int f\, \D^d u,\quad N = \int n\,\D t. \label{eq:nN}
\end{equation}
At a neighbouring point $\vect x+\D\vect x$, the local number of branes is, to first order, advected through the infinitesimal spatial cell of volume $\D^d x$ as \footnote{Crucially, one can show that $\D t'\D^d u' = \D t\D^d u + \mathcal O(\D x^2)$ if one assumes that both external and interaction potentials do not depend on the inverse velocities, which we have assumed throughout.}
\begin{equation}
	\begin{split}
		&f\big(\vect x + \D\vect x,t + \vect u\cdot \D\vect x,\vect u+ (\D\vect x\cdot\vect D)\vect u\big)\,\D t'\,\D^d u' \\
		&= f(\vect x,t,\vect u)\,\D t\,\D^d u + \vect D f\cdot\D \vect x \,\D t\,\D^d u + \mathcal O(\D x^2).
	\end{split}
\end{equation}
In the absence of collisions, the local number of branes in a given inverse-velocity state is advected invariantly; therefore, the first-order term vanishes, resulting in $D_i f = 0$. Otherwise, one can introduce a \textit{collision integral} term $C_i[f]$ to account for the transfer of branes from one inverse-velocity state to another as a result of collisions. By definition,
\begin{equation}
		D_i f = \frac{\partial f}{\partial x^i} + u_i \frac{\partial f}{\partial t} + D_i u_j \frac{\partial f}{\partial u_j} = C_i[f].
	 \label{eq:collisionnal boltzmann carroll}
\end{equation}
The above equation is the Carrollian counterpart of the collisional Boltzmann's equation. 

Let us isolate, around a given point $\vect x$, a brane with inverse velocity $\vect u_1$ and consider an infinitesimal increment $\D^d u_1$ about $\vect u_1$. Owing to direct collisions $\vect{u}_1, \vect{u}_2 \rightarrow \vect{u}'_1, \vect{u}'_2$, for arbitrary $\vect{u}_2$, $\vect{u}'_1$, $\vect{u}'_2$, the density of branes in the $\vect u$-space volume $\D^d u_1$ and along the direction $\partial_i$ decreases by
\begin{equation}
    \D^d u_1 \int \chi_i \,w (\vect{u}_1,\vect{u}_2 ; \vect{u}_1',\vect{u}_2')\, f^{(2)}_{12}\, \D^d u_2\, \D^d u'_1\, \D^d u'_2 , \label{eq:losses}
\end{equation}
where $w(\vect u_1,\vect u_2;\vect u_1',\vect u_2')$ is the \textit{transition rate}, and $f^{(2)}_{12} \coloneqq f^{(2)}(\vect{x}_1, t_1, \vect{u}_1; \vect{x}_2, t_2, \vect{u}_2)$ is the \textit{two-point distribution function}, evaluated in Eq.\ \eqref{eq:losses} at the same spacetime point for both branes, $(\vect x_1,t_1) = (\vect x_2,t_2)$. The former quantifies the probability density for a pair of branes to go from a state $(\vect u_1,\vect u_2)$ to a state $(\vect u_1',\vect u_2')$ upon a binary quasi-local collision. Assuming local conservation of probabilities, it obeys \cite{Pitaevskii1981Physical}
\begin{equation}
    \int \chi_i\,\Big( w (\vect{u}_1, \vect{u}_2; \vect{u}'_1, \vect{u}'_2) - w (\vect{u}'_1, \vect{u}'_2 ; \vect{u}_1, \vect{u}_2)\Big) \,\D^d u'_1\, \D^d u'_2 =0
	\label{eq:unita}
\end{equation}
upon integration over the ``final'' states. The latter is the joint probability of observing the brane $t_1$ in the inverse-velocity state $\vect u_1$ at the point $\vect x_1$, and the brane $t_2$ in the state $\vect u_2$ at the point $\vect x_2$. Moreover, owing to reverse collisions $\vect{u}_1', \vect{u}_2' \rightarrow \vect{u}_1, \vect{u}_2$, the density of branes in the same infinitesimal $\vect u$-space cell increases by
\begin{equation}
    \D^d u_1 \int \chi_i \, w (\vect{u}_1', \vect{u}_2' ; \vect{u}_1, \vect{u}_2) \, f'^{(2)}_{12}\, \D^d u_2 \, \D^d u_1' \, \D^d u_2' , \label{eq:gains}
\end{equation}
with $f'^{(2)}_{12}$ denoting the two-point distribution function for the primed brane states. The balance between gains \eqref{eq:gains} and losses \eqref{eq:losses} then yields the collision integral as
\begin{equation}
	C_i[f] = \int \chi_i\, w (\vect{u}_1', \vect{u}_2' ; \vect{u}_1, \vect{u}_2 ) \left(f'^{(2)}_{12} - f^{(2)}_{12}\right) \D^d u_2\, \D^d u_1'\, \D^d u_2'. \label{eq:explicit Ci}
\end{equation}
The Carrollian collision integral is a vector quantity, as it is aligned with the unit normal vector to the collision plane. Note that the reverse-transition rate has been factorised in Eq.~\eqref{eq:explicit Ci} by virtue of Eq.~\eqref{eq:unita}.

From this explicit formula, we deduce that integrating any conserved quantity against the collision integral yields zero,
\begin{equation}
	\mathcal C(Q) \coloneqq \int Q^i(\vect u)\,C_i[f] \, \D^d u = 0. \label{eq:smearing Qi}
\end{equation}
Indeed, denoting $\D V_{\vect u}\coloneqq \D^d u_1\D^d u_2\D^d u_1'\D^d u_2'$ and $Q^i(\vect u_a)\coloneqq Q^i_a$ to avoid clutter, the above expression may be written as
\begin{subequations}
	\begin{align}
		\mathcal C(Q) &= \int \big(Q^i_1 - Q'^i_1\big)\,\chi_i\,w(\vect u_1',\vect u_2';\vect u_1,\vect u_2)\,f'^{(2)}_{12}\,\D V_{\vect u} \label{eq:16a}\\
		&= \int \big(Q^i_2 - Q'^i_2\big)\,\chi_i\,w(\vect u_1',\vect u_2';\vect u_1,\vect u_2)\,f'^{(2)}_{12}\,\D V_{\vect u}.\label{eq:16b}
	\end{align}
\end{subequations}
Eq.\ \eqref{eq:16a} is obtained by substituting $C_i[f]$, as given in Eq.\ \eqref{eq:explicit Ci}, into Eq.\ \eqref{eq:smearing Qi} and using the property \eqref{eq:unita}, while Eq.\ \eqref{eq:16b} is simply a rewriting of Eq.\ \eqref{eq:16a}, owing to the fact that swapping the labels of the colliding branes does not affect the transition rate. Adding these two equivalent expressions then allows us to conclude that $\mathcal C(Q)$ vanishes when
\begin{equation}
	\big(Q^i_1+Q^i_2\big)\chi_i = \big(Q'^i_1+Q'^i_2\big)\chi_i. \label{eq:conserved upon coll}
\end{equation}
We show in Appendix \ref{app:conservation} that this identity is satisfied by any quantity conserved under a binary brane collision, hence the stated result.

\section{Carrollian fluid equations}
\label{sec:fluid}

We now seek to establish the conservation equations for a fluid as the statistical manifestation of the microphysical quantities conserved during collisions. We pass from microphysical to statistical quantities by averaging over inverse velocities. For any quantity $Q = Q(\vect x,t,\vect u)$, we define its {local average} as
\begin{equation}
\langle Q\rangle(\vect x,t) \coloneqq \frac{1}{n}\int Q\,f\,\D^d u,
\end{equation}
where the normalisation factor $1/n$ is chosen so that $Q$ and its average $\langle Q\rangle$ have the same units. The local {density} of the macroscopic fluid is $\rho(\vect x,t) \coloneqq \sigma\, n$. Computing the first moment of the distribution function yields a notion of collective motion of the brane perturbations, which we call the {fluid inverse velocity}, $\beta^i(\vect x,t) \coloneqq \langle u^i\rangle$. Since Eq.\ \eqref{eq:smearing Qi} holds for any quantity $Q^i$ conserved in a collision, averaging the Boltzmann--Carroll equation, Eq.\ \eqref{eq:collisionnal boltzmann carroll}, contracted with $Q^i$, yields the following transport equation:
\begin{equation}
\partial_i \big(n\langle Q^i\rangle\big) + \partial_t \big(n\langle u_i Q^i\rangle\big) - n\langle D_i Q^i\rangle = 0. \label{eq:genfluid}
\end{equation}
To obtain this, we have discarded a boundary term arising from integration by parts over inverse velocities, assuming that the distribution function is localised in momentum space.\footnote{Note that Eq.\ \eqref{eq:genfluid} holds if and only if the external power does not depend on inverse velocities. Furthermore, the derivation also requires to note that $\frac{\partial}{\partial u_j}$ annihilates the trace-free part of $D_i u_j$ on shell. Indeed, prior to imposing equations of motion, jet coordinates $u_i$ and $D_i u_j$ are independent. Going on shell amounts to impose a constraint between $u_i$'s and their total divergences $D_i u^i$, but the trace-free components are still independent of $u_i$.}

The transport equation \eqref{eq:genfluid} must now be specified for the relevant conserved quantities in order to derive the corresponding fluid equations. The first interesting conserved quantity upon collision is the energy current, $Q^i(\vect u)=E^i(\vect u)$ from Eq.\ \eqref{eq:energy current}. In the presence of external power $P$, it obeys $D_i E^i = P$ on shell. Averaging the Carroll--Boltzmann equation thus provides the scalar fluid equation
\begin{equation}
	\mathscr E \coloneqq \partial_i(\rho\beta^i) + \partial_t\big(\eta + \rho\bm\upbeta^2\big) - \frac{\rho}{\sigma} P = 0, \label{eq:carr E}
\end{equation}
where $\eta \coloneqq \rho \langle (\vect u-\bm\upbeta)^2\rangle$ is the local fluid energy. We shall refer to this as the \textit{Carroll energy equation}. Next, we consider the kinetic stress tensor $Q^i{}_j(\vect u) = K^i{}_j(\vect u)$ given by Eq.\ \eqref{eq:stress tensor}. On shell, it obeys $D_i K^i{}_j = u_j P$ when external power is exerted on the system, which can be viewed as the Carrollian instance of the \textit{theorem of kinetic energy}. Inserting this in Eq.\ \eqref{eq:genfluid} yields the \textit{Carroll momentum equation},
\begin{equation}
	\begin{split}
		\mathscr G_j &\coloneqq \partial_i\big(\mathcal P^i{}_j + \rho \beta^i\beta_j -\tfrac12\delta^i{}_j\rho\bm\upbeta^2\big) \\
		&\hspace{.5cm} + \partial_t \big(\pi_j + \beta_j\eta + \beta_i \mathcal P^i{}_j +\tfrac12 \rho\beta_j\bm\upbeta^2\big) - \frac{\rho}{\sigma}\beta_j P = 0,
	\end{split}\label{eq:carr Pj}
\end{equation}
where the internal pressure tensor $\mathcal P^i{}_j$ and heat current $\pi_j$ of the fluid are defined by
\begin{subequations}
	\begin{align}
		\mathcal P^i{}_j &\coloneqq \rho \big\langle (u^i-\beta^i)(u_j-\beta_j) - \tfrac12 \delta^i{}_j(\vect u-\bm \upbeta)^2\big\rangle, \label{eq:Pij} \\
		\pi_j &\coloneqq \tfrac12 \rho \big\langle (u_j-\beta_j)(\vect u-\bm\upbeta)^2\big\rangle.
	\end{align}
\end{subequations}
As measured in the fluid ``rest frame,'' by which we mean averaged relative to the fluid inverse velocity $\bm\upbeta$, the spacetime fields $\eta$, $\mathcal P^i{}_j$ and $\pi_j$ are intrinsic to the fluid and therefore invariant under Carroll boosts. In the Galilean case, a quantity that is trivially conserved in collisions is the mass of the colliding molecules, which leads to the continuity equation upon averaging. Here, this role is played by $\sigma$, or more specifically the local strain tensor $Q^i{}_j = \sigma \delta^i{}_j$. Local strain conservation then yields the Carrollian analogue of the \textit{continuity equation}:
\begin{equation}
		\mathscr H_i \coloneqq \partial_i\rho + \partial_t(\rho\beta_i) = 0. \label{eq:carr Ci}
\end{equation} 
Remarkably, Eqs. \eqref{eq:carr E}, \eqref{eq:carr Pj} and \eqref{eq:carr Ci} correspond exactly to the Carroll fluid equations derived from the vanishing-speed-of-light limit of special-relativistic conservation laws of the form $\partial_\mu T^\mu{}_\nu=0$, upon appropriately expanding the relativistic quantities in powers of the speed of light \cite{Ciambelli:2018xat,Petkou:2022bmz}, and up to the incorporation of an external-power term. More details on this correspondence are provided in Appendix~\ref{app:matching}.

Unlike in the Galilean case, we need a supplementary equation to ensure that Carroll fluid dynamics is invariant under the full Carroll group. This equation does not follow from a Noether invariant, but rather from a kinematic constraint. Indeed, since the inverse velocity is a gradient, it is curl-free, $D_{[i}u_{j]}=0$. As this property is preserved under collisions, there exists a conserved quantity $Q^i{}_{jk}(\vect u) = Q^i{}_{[jk]}(\vect u) = 2\sigma \delta^i{}_{[j}u_{k]}$, which is identically conserved, \textit{i.e.}\ $D_i Q^i{}_{jk}(\vect u)=0$ without invoking the equations of motion. Substituting this additional conserved current into the averaged Carroll--Boltzmann equation gives
\begin{equation}
\mathscr I_{ij} \coloneqq 2\partial_{[i}\big(\rho\beta_{j]}\big) =0. \label{eq:carr Iij}
\end{equation}
We refer to \eqref{eq:carr Iij} as the \textit{irrotationality equation}, since it reflects the fact that flows induced by dynamical Carroll space-filling branes are necessarily irrotational. Furthermore, averaging the Carroll--Boltzmann equation against the rotation and boost charges does not impose any further non-trivial constraints: rotation invariance yields $\bar T_{[ij]} = 0$, while, as in the Galilean case, boost invariance equates the energy flux $\rho\beta^i$ appearing in the spatial divergence in Eq.\ \eqref{eq:carr E} with the matter current $\rho\beta_i$ appearing in the time derivative in Eq.\ \eqref{eq:carr Ci}. Finally, it can be shown that the system of Eqs. \eqref{eq:carr E}, \eqref{eq:carr Pj}, \eqref{eq:carr Ci} and \eqref{eq:carr Iij} transforms tensorially under Carroll transformations. In particular, we have
\begin{equation}
\begin{split}
\mathscr H_i &\mapsto \mathscr H_i,\quad
\mathscr E \mapsto \mathscr E - b^i\mathscr H_i,\quad \mathscr I_{ij} \mapsto \mathscr I_{ij} + 2 b_{[i}\mathscr H_{j]}. \\
\mathscr G_i &\mapsto \mathscr G_i - b_i\mathscr E + b_i b^j\mathscr H_j - \tfrac{1}{2}\vect b^2\mathscr H_i + \mathscr I_{ij}b^j,
\end{split}
\end{equation}
under Carroll boosts, which act as $(u^i,\beta^i) \mapsto (u^i-b^i,\beta^i-b^i)$. Clearly, the supplementary equation $\mathscr I_{ij}=0$ is necessary to ensure Carroll-boost invariance of the momentum equation.

\section{Elements of Carroll thermodynamics}
\label{sec:thermo}
Having established a microscopic interpretation of the Carrollian fluid equations, we now present the first elements of a theory of Carrollian thermodynamics. In Galilean physics, the dynamical laws are invariant under time reversal. However, irreversibility emerges at the statistical level because of the molecular-chaos hypothesis (``\textit{Sto\ss zahlansatz}''), which states that the two-point distribution function factorises before collision. This in turn allows a quantity, Boltzmann's entropy $S(t)$, to increase over time, a result referred to as the $H$-theorem. In Carrollian physics, since the natural microscopic object is extended and there is no notion of ``before'' or ``after'' a collision, there is no clear equivalent of the molecular-chaos hypothesis and therefore no clear analogue of the $H$-theorem. Yet, we can still introduce entropy density $s$ and entropy current $\vect j$ as
\begin{equation}
	s\coloneqq -k_\textrm{B}\int f\ln f\,\D^d u,\quad \vect j\coloneqq -k_\textrm{B}\int\vect u \, f \ln f\,\D^d u,
\end{equation}
where $k_\textrm{B}$ is Boltzmann constant. They obey the following transport equation
\begin{equation}
	\partial_i s + \partial_t j_i = -k_\textrm{B}\int \ln f\, C_i[f]\,\D^d u, \label{eq:transport s}
\end{equation}
on account of Eq.\ \eqref{eq:collisionnal boltzmann carroll}. Equilibrium is then defined as the situation in which the right-hand side of Eq.\ \eqref{eq:transport s} vanishes. All this echoes preliminary observations based on the two-dimensional Galilei--Carroll duality properties highlighted in \cite{Athanasiou:2024lzr}.

Let us recall that microstates of a Galilean system populate energy levels, labeled by a scalar constant $E$ that is conserved in time by virtue of the equations of motion. In order to define the concept of temperature, one assumes that the finite-size system $\mathscr S_1$ under scrutiny is in contact with an external thermostat $\mathscr S_2$ composed of infinitely many degrees of freedom, for it to be insensitive to fluctuations. At equilibrium, the thermostat, which has a fixed temperature, imposes the latter to the system: this defines the canonical ensemble, as reviewed in Appendix \ref{app:canonical}. It translates in our Carrollian setup as follows. First, the dynamical statement of time conservation of $E$ is traded for the on-shell conservation of the stress tensor \eqref{eq:stress tensor} over space, \textit{i.e.}\ $D_i T{}^i{}_j = 0$. Unlike its Galilean counterpart $\dot E = 0$, this equations allows for space-dependent solutions: we shall therefore define a Carroll microstate at point $\vect x$ as the collection of branes with on-shell stress tensor $T^{ij}(\vect x)$ at this point. Subsequently, a Carrollian thermostat is an external medium with infinite extension in the $t$-direction imposing its spaceture to the system, in the sense that
\begin{equation}
	\frac{\partial S_1}{\partial T_1^{ij}(\vect x)} \stackrel{!}{=} \frac{\partial S_2}{\partial T_2^{ij}(\vect x)} \coloneqq \alpha_{ij}(\vect x),\ \forall\vect x. \label{eq:equilcarr}
\end{equation}
This equation, which can be tamed as the zeroth principle of Carrollian thermodynamics, is proven using the microcanonical approach within our brane setup in Appendix \ref{app:canonical}. The symmetric rank-two tensor field $\bm\upalpha = \alpha_{ij}(\vect x)\D x^i\D x^j$ plays the role of inverse temperature: we therefore refer to its inverse, $\mathcal S_{ij}\coloneqq\alpha_{ij}^{-1}/k_\textrm{B}$, as the \textit{Carrollian local spaceture}.\footnote{The word ``\textit{temperature}'' has a common etymologic origin with the word \textit{time}, and Galilean energy is associated to time translation. Here the Carrollian stress tensor is associated with space translation and the ``evolution'' is over space: we therefore propose to dub ``\textit{spaceture}'' the Carrollian analogue of temperature.}

The equilibrium distribution $f_0$, which we shall refer to as the \textit{Carroll--Maxwell--Boltzmann distribution}, is obtained by maximising the total entropy $S(\vect x) = \int s(t,\vect x)\,\D t$ of the system and the thermostat under the constraints \eqref{eq:nN} and \eqref{eq:equilcarr}. Neglecting interactions to first order, we find
\begin{equation}
	f_0(\vect x,\vect u) \coloneqq \frac{1}{Z_c(\bm\upalpha(\vect x))}\exp\left(-\alpha_{ij}(\vect x) K^{ij}\big(\vect u-\bm\upbeta(\vect x)\big)\right), \label{eq:maxwellboltz}
\end{equation}
where velocities are reported to the local average inverse velocity $\bm\upbeta(\vect x)$ of the system. It is further required that $\tilde\alpha_{ij}\coloneqq 2\alpha_{ij} - \alpha^k{}_k\delta_{ij}$ be positive-definite, in order for the integral of $f_0$ over $\vect u$ to be convergent. This mirrors the positivity of Galilean temperature and can be assumed in the Carrollian instance for any $d\neq 2$. The two-dimensional case calls for a separate treatment, which we will not undertake in the present letter. The argument of the exponential function can be re-expressed as $-\tfrac12 \tilde\alpha^{ij}\sigma(u_i-\beta_i)(u_j-\beta_j)$, which manifestly displays its Gaussian character. Finally, the normalisation prefactor
\begin{equation}
	Z_c(\bm\upalpha(\vect x)) = \frac{1}{n(\vect x)}\int \exp\left(-\alpha_{ij}(\vect x) K^{ij}\big(\vect u-\bm\upbeta(\vect x)\big)\right) \D^d u = \frac{\sqrt{\det\tilde\alpha(\vect x)^{-1}}}{n(\vect x)} \left(\frac{2\pi}{\sigma}\right)^{\frac{d}{2}}, \label{eq:normalisation CMB}
\end{equation}
fixed by the requirement that $f_0$ integrates to $n$ in Eq. \eqref{eq:nN}, provides a well-defined notion of canonical partition function in the Carrollian regime. 

Having defined such an equilibrium distribution allows for multiple things, among which the settlement of the Carroll concept of perfect fluid. As in Galilean physics, we characterise here a perfect fluid by a distribution function approximated by $f_0$ in the local reference frame of the fluid, with spacetime dependent parameters $n$, $\bm\upbeta$ and $\alpha_{ij}$. It is then a simple matter of algebra to obtain the specific energy $\eta$, heat current $\pi_i$ and pressure tensor $\mathcal P_{ij}$ for a Carrollian perfect fluid as \footnote{Concretely, perfect Carrollian fluids have been defined in \cite{Ciambelli:2018xat} by performing the Carrollian limit of perfect relativistic fluids whose pression tensor is isotropic. The limit does not change this matter of fact, hence the resulting Carrollian fluid obeys $\pi^i = 0$, as we find, but also $\mathcal P_{ij}\propto \delta_{ij}$. However, from our purely Carrollian analysis, since equilibrium can only be imposed pointwise in space, we have not found any evidence to interpret deviations from isotropy in $\mathcal P_{ij}$ as the manifestation of dissipative processes. Hence, perfect-fluidic $\mathcal P_{ij}$ is not constrained to be isotropic but merely obeys Eq.\ \eqref{eq:perfect}.}
\begin{equation}
		\eta = k_\textrm{B}\,n\,\tilde{\mathcal S}{}^i{}_i,\quad \pi_i=0,\quad \mathcal P_{ij} = k_\textrm{B}\,n\,\big(\tilde{\mathcal S}_{ij} - \tfrac12\delta_{ij}\tilde{\mathcal S}{}^k{}_k\big), \label{eq:perfect}
\end{equation}
where $\tilde{\mathcal S}_{ij}\coloneqq \tilde\alpha_{ij}^{-1}/k_{\textrm B}$ is the Gaussian trace-adjusted spaceture. It agrees with the heuristic definition of \cite{Ciambelli:2018xat}. Let us make a last remark. In Galilean physics, temperature is associated with the particles' energetic excitation. Here, the spaceture is associated with the excitation of the branes' stress tensor: the more the branes wiggle, the more readily they can collide in the manner envisaged here, and the higher the spaceture. Since space-filling branes are extended objects that can bend in all spatial directions, the physical quantity that measures this effect must be tensorial. This nicely explains why the spaceture is a rank-two spacelike tensor rather than a scalar, as in the usual case.

\section{Discussion}
Concluding, let us now summarise our results and outline a few future directions. In this work, we have offered a first-principles derivation of the Carrollian fluid equations, Eqs. \eqref{eq:carr E}--\eqref{eq:carr Ci}, from a microscopic analysis in terms of interacting instantonic space-filling branes, as introduced in \cite{Ciambelli:2018xat}, and have thereby laid the first foundations of a theory of Carrollian thermodynamics. Our derivation provides a novel interpretation of Carrollian dynamics and calls for many concrete applications in the future.

Two natural extensions of our analysis are the investigation of Carrollian quantum mechanics through the quantisation of systems of space-filling branes, and the extension to conformally-symmetric system in two and three dimensions, which are relevant for flat-space holography. In particular, it would be interesting to compare the resulting definition of entropy with the computation of gravitational entropy of three-dimensional flat cosmologies from a boundary computation \cite{Barnich:2012xq,Bagchi:2014iea,Basu:2015evh}, and to look for possible four-dimensional extensions of this matching involving realistic black hole solutions. Furthermore, as Carroll geometry arises at black hole horizons \cite{Penna:2017bdn,Penna:2018gfx,Chandrasekaran:2018aop,Ciambelli:2019lap,Chandrasekaran:2021hxc,Freidel:2022vjq,Redondo-Yuste:2022czg,Ciambelli:2023mir,Freidel:2024emv,Husnugil:2025edm} (see also \cite{jankiewicz1954espaces,Vogel1965,Datcourt1967,Henneaux:1979vn}), it would be interesting to determine whether our definition of Carroll entropy is able to shed light on the fact that black hole entropy scales with the area rather than the volume, as a by-product of Carrollian dynamics at the horizon \cite{Bekenstein:1973ur,Bekenstein:1974ax}. We leave all these exciting perspectives for future endeavours.

\section*{Acknowledgements}
We thank Mathieu Beauvillain, Daniel Grumiller, Cédric Lorcé, Simon Pekar and Antony Wendels for useful discussions. The work of AF is partially supported by the Fonds de la Recherche Scientifique -- FNRS Belgium (convention IISN 4.4503.15), as well as by research funds from the Solvay Family. The work of MV is supported by the Fonds de la Recherche Scientifique -- FNRS under the Grant No. T.0047.24.

\appendix
\setcounter{secnumdepth}{1}

\section{Carrollian units}
\label{app:carrvsgal}
In this Appendix, we discuss how physical units of the various quantities introduced in the main text compare between Galilean and Carrollian mechanics. As a starting point, we require that Galilean and Carrollian actions, $S_{\text G} = \int L_{\text G}\D t$ and $S_{\text C} = \int L_{\text C}\D^d x$, have the same units, namely energy $\times$ time. Therefore, the tension $\sigma$ introduced in Eq.\ \eqref{eq:action carr} has units of 
\begin{equation}
	\Sigma \coloneqq [\sigma] = \text L^{4-d}\, \text T^{-3}\, \text M.
\end{equation}
In the Carrollian setting, it is more economic to trade the usual mass unit $\text M$ for $\Sigma$ in order for the usual dynamical quantities to have units that do not depend on the number of dimensions. To convert Galilean units into Carrollian units, we follow the general rule
\begin{equation}
	\text M\leftrightarrow\Sigma,\quad \text L\leftrightarrow \text T. \label{eq:rule units}
\end{equation}
As a concrete example, while the Galilean Lagrangian $L_{\text G}$ has units of mass $\times$ squared velocity, $[L_{\text G}] = \text M\,\text L^2\,\text T^{-2}$, the Carrollian Lagrangian $L_{\text C}$ has units of tension $\times$ squared inverse velocity, $[L_{\text C}] = \Sigma \, \text T^2\, \text L^{-2}$. 

In the Galilean context, mechanical Lagrangian functions like $L_{\text G}$ are commensurable to an energy. In the Carrollian case, we decide to keep calling ``energy'' the Noether charge that is conserved by virtue of time-translation invariance. It takes the form of a spatial current $E^i = \sigma u^i$, whose units are $\Sigma\, \text T\, \text L^{-1}$. The equation of motion for the inverse-velocity field $\vect u$ reads $D_i E^i = P$ where $P$ has units of Carrollian energy current over lenght. Since Carrollian evolution occurs in space, this accurately reflects the concept of power in Galilean mechanics, which is an energy flow over time, and we therefore refer to $P$ as Carrollian power, with units $\Sigma\, \text T\, \text L^{-2}$. 

Furthermore, space-translation invariance in the Galilean context leads to momentum conservation, $\dot{\vect p} = \bm 0$ on shell. In the presence of external forces $\vect F$, one finds instead $\dot{\vect p} = \vect F$, which is coined as the fundamental principle of Galilean dynamics. Moving to Carrollian realm, space-translation invariance implies conservation of the stress tensor, $D_i T^i{}_j=0$ on shell. The latter has units of Carrollian energy current $\times$ inverse velocity: $[T^{ij}] = \Sigma\, \text T^2\,\text L^{-2}$. If $L_{\text C}$ depends explicitly on $\vect x$, the conservation law is modified as $D_i T^i{}_j = F_j$ where $F_j = \partial L_{\text C}/\partial x^j$ denote components of the external Carrollian force, with units of $\Sigma\, \text T^2\,\text L^{-3}$. Expectedly, the rule \eqref{eq:rule units} maps units of force onto units of power and reciprocally. The transcription between Galilean and Carrollian fluid quantities follows from the above discussion, and agrees, in particular, with \cite{Athanasiou:2024lzr}.

\section{Conservation laws upon collision}
\label{app:conservation}

Here, we study the kinematics of a binary brane collision. As far as conserved quantities are concerned, the analysis has been done in the main text. Indeed, binary collisions are supposed to occur locally in time and along a codimension-one surface in space so that the action of any external power can be neglected; this is in line with Boltzmann's analysis. Therefore, conserved quantities are still given by Eqs. \eqref{eq:energy current} and \eqref{eq:stress tensor} for $a=1,2$ only. We generically denote as $D_i Q^i=0$ the related conservation equation. Since the conserved quantities are additive, $Q^i = Q^i(\vect u_1) + Q^i(\vect u_2)$ on one side of the collision and $Q'^i = Q^i(\vect u_1') + Q^i(\vect u_2')$ on the other side of the collision. At any point $P\in\mathcal I$, see Fig.~\ref{fig: collision between two branes}, we can decompose $Q^i = Q_0 \chi^i + Q^i_\bot$ where $\chi_i Q^i_\bot = 0$ by definition, so that the conservation equation can be rewritten as $\bm\upchi[Q_0] + D_i Q^i_\bot = 0$. Because of our hypothesis of locality, the inverse-velocity field $\vect u$ may experience a discontinuity in the $\bm\upchi$-direction but is required to be smooth in any orthogonal direction. In terms of the conserved quantities, this translates into the following conditions
\begin{equation}
	\bm\upchi[Q_0] = \frac{Q'_0 - Q_0}{\delta x} + \mathcal O\!\left((\delta x)^0\right),\quad D_i Q^i_\bot = \mathcal O\!\left((\delta x)^0\right).
\end{equation}
where $\delta x$ is a small increment in the direction of $\bm\upchi$ in the vicinity of $P$. Using the conservation equation, we finally obtain
\begin{equation}
	Q'_0 - Q_0 = \delta x\,\left(\bm\upchi[Q_0] + \mathcal O\!\left((\delta x)^0\right)\right) = \mathcal O(\delta x).
\end{equation}
Taking the $\delta x\to 0$ limit expectedly yields Eq.\ \eqref{eq:conserved upon coll}.

\section{Carrollian fluids from non-Lorentzian limits}
\label{app:matching}
In \cite{Ciambelli:2018xat,Ciambelli:2018wre,Petkou:2022bmz}, a complete analysis of the two complementary non-Lorentzian limits of the relativistic fluid equations $\nabla_\mu T^\mu{}_\nu = 0$ has been performed on generally-covariant backgrounds, using either Zermelo or Papapetrou--Randers coordinate systems (see also \cite{deBoer:2017ing,deBoer:2021jej,deBoer:2023fnj} for complementary analyses). We quote here some results from the study of the Carrollian, \textit{i.e.}\ vanishing-speed-of-light limit, applied to flat backgrounds, and show that the fluid equations obtained in this limit match with the predictions of the kinetic theory we developed in the main text.

First, we review the hypotheses on the small-$c$ expansion of the various relevant fields, where $c$ denotes the speed of light. We consider $(d+1)$-dimensional Minkowski spacetime, covered by Cartesian coordinates $x^\mu = (t,\vect x)$ and endowed with the flat metric $g_{\mu\nu} = \text{diag}(-c^2,1,\dots,1)$. The fluid velocity is a given timelike congruence $u^\mu$, normalised as $g_{\mu\nu}u^\mu u^\nu = -c^2$. In the $c\to 0$ limit, it collapses into the unique field of observers $\bfupsilon$ while the metric collapses to the degenerate Carrollian metric $g_{\mu\nu} = \text{diag}(0,1,\dots,1)$. The spacelike projection of $u^\mu$ is therefore assumed to vanish at zero $c$ but there should remain a kinematic parameter in the limit for the latter to be non-trivial. One then posits $u^t = 1+\mathcal O(c^2)$ and $u^i = c^2\beta^i + \mathcal O(c^4)$, where $\beta^i$ has units of inverse velocity, to prepare the Carroll limit. 

The energy-momentum tensor $T^{\mu\nu}$ of the fluid can be decomposed in longitudinal and transverse components as
\begin{equation}
	T^{\mu\nu} = (\varepsilon+p)\frac{u^\mu u^\nu}{c^2} + p g^{\mu\nu} + \tau^{\mu\nu} + \frac{2u^{(\mu} q^{\nu)}}{c^2},
\end{equation}
where $\varepsilon$ is the energy density, $p$ the pressure density, $q^\mu$ the heat current and $\tau^{\mu\nu}$ the viscous stress tensor. By definition, the two latter objects are purely transverse tensors with respect to the congruence $u^\mu$, \textit{i.e.}\ $u_\mu\tau^{\mu\nu} = 0$ and $u_\mu q^\mu = 0$. The functions $\varepsilon$, $p$ and $q^\mu$ are specific to the fluid and expanded as
\begin{equation}
	\begin{split}
		\varepsilon &= \eta+\mathcal O(c^2),\quad p = \varpi + \mathcal O(c^2), \\
		q^i &= c^2\pi^i+\mathcal O(c^4),\quad \tau^{ij} = -\frac{\Sigma^{ij}}{c^2} - \Xi^{ij} + \mathcal O(c^2),
	\end{split}
\end{equation}
by assumption in the $c\to 0$ limit, and all the coefficients in the above expansions are now independent of the speed of light $c$. 

On account of these expansions, conservation laws involving the relativistic fluid energy--momentum tensor, in the absence of external forces, separate into three independent equations in the limit
 \begin{equation}
	-\partial_\mu T^\mu{}_t = \mathscr E + \mathcal O(c^2),\quad \partial_\mu T^\mu{}_i = \frac{\mathscr H_i}{c^2} + \mathscr G_i + \mathcal O(c^2).
\end{equation}
One recognises the Carroll energy, continuity and momentum equations, written as
\begin{subequations}
	\begin{align}
		\mathscr E &= \partial_t\Pi + \partial_i\Pi^i = 0, \label{eq:EMarios}\\
		\mathscr H_i &= \partial_t\Pi_i + \partial_j \tilde\Pi^j{}_i = 0, \label{eq:CiMarios} \\
		\mathscr G_j &= \partial_t P_j + \partial_i\Pi^i{}_j = 0, \label{eq:PjMarios}
	\end{align} \label{eq:Marios}
\end{subequations}
in terms of the \textit{Carrollian momenta} \cite{Ciambelli:2018xat,Ciambelli:2018ojf,Petkou:2022bmz}
\begin{equation}
	\begin{split}
		\Pi &= \eta - \Sigma^{ij}\beta_i\beta_j,\quad \Pi^i = -\Sigma^{ij}\beta_j, \\
		\Pi^{ij} &= \varpi\delta^{ij} - \Xi^{ij}, \quad \tilde\Pi{}^{ij} = -\Sigma^{ij}, \\
		P_j &= \pi_j + \eta\beta_j + \beta^i(\varpi\delta_{ij}-\Xi_{ij}).
	\end{split}
\end{equation}
Note that the irrotationality equation \eqref{eq:carr Iij} has no correspondence in the Carrollian-limit framework, as it is specific to the microphysical description in terms of space-filling branes. Eqs. \eqref{eq:Marios} then correspond to Eqs. \eqref{eq:carr Ci}--\eqref{eq:carr Pj} if one identifies
\begin{equation}
	\Sigma^{ij} \equiv - \rho\delta^{ij},\quad \varpi\delta^{ij}-\Xi^{ij} \equiv P^{ij} + \rho\beta^i\beta^j - \tfrac12\rho\bm\upbeta^2\delta^{ij}.
\end{equation}
These identifications agree with \cite{Athanasiou:2024lzr}, where two-dimensional Carrollian fluids have been discussed from a more mathematical perspective.

\section{Microcanonical ensemble}
\label{app:canonical}

In this Appendix, we discuss the definition of the microcanonical ensemble in our Carrollian setup and justify our definition of spaceture given in the main text. For completeness, we start by offering a brief review on the Galilean case.

\subsubsection*{The Galilean case}

In Galilean physics, the central property of a thermodynamically isolated system is energy conservation, which entails that the total mechanical energy of such a system is a constant real number. The existence of such a constant lies at the foundation of Galilean statistical physics. The microcanonical approach amounts to considering an isolated system $\mathscr S$ of fixed energy $E$. Let $\{ m\}$ denote the set of microstates that can be occupied by the system, and $p_m$ the probability for the system to be in the microstate $m$. We assume for simplicity that $\{ m\}$ is a (possibly infinite-dimensional) countable set; the passage to continuous spectra is straightforward, at least within the scope of our analysis. One conventionally denotes by $W(E)$ the number of microstates with energy $E$. The Gibbs entropy of the system is defined as
\begin{equation}
	S(\{p_m\}) \coloneqq -k_{\textrm B}\sum_m p_m \ln p_m,
\end{equation}
where the discrete sum is taken over all accessible microstates. The equilibrium configuration is reached when, by definition, the latter is maximal. The objective function to be extremised is then
\begin{equation}
	\frac{1}{k_{\textrm B}}G(\{p_m\},\lambda) \coloneqq -\sum_m p_m \ln p_m - \lambda\left(\sum_m p_m - 1\right), \label{eq:obj gal}
\end{equation}
where $\lambda\in\mathbb R$ is a Lagrange multiplier imposing the completeness relation, according to which all probable configurations must define a partition of the identity. The function \eqref{eq:obj gal} reaches its maximum if and only if all probabilities are equal, $p_m^* = W(E)^{-1}$. The Gibbs entropy of this equilibrium state is given by Boltzmann's famous formula:
\begin{equation}
	S(\{p_m^*\}) = -k_{\textrm B}\sum_m p_m^* \ln p_m^* = k_{\textrm B} \ln W(E). \label{eq:boltzmann formula}
\end{equation} 
In order to define a notion of temperature, we now consider two systems $\mathscr S_1$ and $\mathscr S_2$ that can exchange energy. The total system $\mathscr S = \mathscr S_1\cup \mathscr S_2$ is isolated, with constant total mechanical energy $E$, and can therefore be analysed through a microcanonical approach. The energies $E_1$ and $E_2$ of the coupled systems always sum to $E=E_1+E_2$. We still denote by $W_1(E_1)$, respectively $W_2(E_2)$, the number of microstates of the system $\mathscr S_1$ of energy $E_1$, respectively $\mathscr S_2$ of energy $E_2$. The probability of finding the system $\mathscr S_1$ in any microstate of energy $E_1$ is given by the product of the probability of finding this system in \textit{a} given microstate $m$ with energy $E_1$, times the degeneracy of all possible microstates with energy $E_1$, \textit{i.e.},
\begin{equation}
	p(E_1) = \frac{W_2(E-E_1)}{W(E)}W_1(E_1). \label{eq:pE1}
\end{equation}
Again, by normalisation of probabilities, we have
\begin{equation}
	\sum_{E_1}p(E_1) = 1\qquad\Rightarrow\qquad W(E) = \sum_{E_1} W_1(E_1)W_2(E-E_1).
\end{equation}
Combining both pieces of information and using Eq. \eqref{eq:boltzmann formula}, we obtain
\begin{equation}
	p(E_1) = \frac{1}{W(E)}\exp \left(\frac{S(E_1)}{k_{\textrm B}}\right),\qquad S(E_1) \coloneqq S_1(E_1) + S_2(E-E_1), \label{eq:d6}
\end{equation}
where $S_1(E_1)$, respectively $S_2(E_2)$, denotes the entropy of $\mathscr S_1$ at energy $E_1$, respectively $\mathscr S_2$ at energy $E_2$. At equilibrium, \textit{i.e.} for $E_1=E_1^*$, the total entropy is assumed to reach a maximum, which yields
\begin{equation}
	\left.\frac{\partial S}{\partial E_1}\right|_{E_1^*} = 0 \qquad\Rightarrow\qquad \left.\frac{\partial S_1}{\partial E_1}\right|_{E_1^*} = \left.\frac{\partial S_2}{\partial E_2}\right|_{E-E_1^*}. \label{eq:equal of temp}
\end{equation}
Equilibrium configurations are then characterised by the fact that both systems share a thermodynamical quantity, given by
\begin{equation}
	\frac{1}{T}\coloneqq \frac{\partial S}{\partial E}, \label{eq:micro temp}
\end{equation}
where $T$ is the microcanonical notion of temperature. The equality \eqref{eq:equal of temp} is often referred to as the zeroth principle of thermodynamics. 

This naturally leads us to the notion of a Galilean thermostat. Indeed, if we suppose that the second system, $\mathscr S_2$, is ``much bigger'' than the first one, $\mathscr S_1$, in the sense that $\text{card}(\{m_2\})\gg \text{card}(\{m_1\})$, the respective energies of the coupled systems obey $E_1 \ll E_2$, and therefore $E_2 = E-E_1\simeq E$. Thus, the entropy of the second system can be approximated to first order in the deviations of $E_2$ around $E$ as
\begin{equation}
	S_2(E_2) = S_2(E) - E_1\left.\frac{\partial S_2}{\partial E_2}\right|_{E_2=E} + \mathcal O(E_1^2) = S_2(E) - \frac{E_1}{T_2} + \mathcal O(E_1^2),
\end{equation}
owing to Eq. \eqref{eq:micro temp}. Using again Boltzmann's formula \eqref{eq:boltzmann formula} for the Gibbs microcanonical entropy, the previous equation is immediately rewritten as
\begin{equation}
	W_2(E-E_1) = W_2(E)\exp\left(-\frac{E_1}{k_{\textrm B}T_2}\right) \label{eq:d10}
\end{equation}
up to negligible terms, consistently with the approximation made above. The probability for $\mathscr S_1$ to be found in a given microstate $m$ with energy $E_1 = E_1^m$ is then computed from Eqs. \eqref{eq:pE1} and \eqref{eq:d10} as
\begin{equation}
	p_m(E_1^m) = \frac{p(E_1^m)}{W_1(E_1^m)} = \frac{W_2(E-E_1^m)}{W(E)} = \frac{1}{Z_c(T_2)}\exp\left(-\frac{E_1^m}{k_{\textrm B}T_2}\right), \label{eq:mb}
\end{equation}
where the normalisation constant $Z_c$ is a function of the equilibrium temperature only:
\begin{equation}
	Z_c(T) \coloneqq \sum_m \exp \left(-\frac{E_1^m}{k_{\textrm B}T}\right).
\end{equation}
It is fixed such that $\sum_m p_m(E_1^m) = 1$. Equation \eqref{eq:mb} displays the discrete version of Maxwell--Boltzmann distribution. As a result, the second system $\mathscr S_2$ imposes its own temperature $T_2$ on the system $\mathscr S_1$ and is therefore seen as a thermostat at temperature $T_2$. Indeed, the fact that $T_2$ is stable under exchanges of energy at the interface separating $\mathscr S_1$ and $\mathscr S_2$ is related to the resilience of the second system to fluctuations of temperature, which is fundamentally encoded in the fact that the second system is, morally speaking, infinite with respect to the first one. In passing, $Z_c(T)$ encodes the equilibrium partition function in the canonical ensemble, which consists of a system $\mathscr S_1$ in thermal equilibrium with a thermostat $\mathscr S_2$ at temperature $T$.

\subsubsection*{The Carrollian case}

We now turn our attention to the Carrollian microcanonical ensemble. As a first and instrumental observation, one shall note that Galilean energy conservation is now replaced by the on-shell conservation of the stress tensor, $D_i T^{ij} = 0$. Unlike $\dot E=0$, this equation admits $\vect x$-dependent solutions, which forces us to consider equilibrium conditions pointwise in space. This matter of fact actually reflects that evolution on Carroll manifolds should better be understood over space, which is generally multi-dimensional, after the lightcones have shrunk. The related microcanonical ensemble can be constructed as follows for any spatial dimensions $d\neq 2$.\footnote{In $d=2$, the traceless character of the kinetic stress tensor, fundamentally rooted to conformal invariance, imposes a refinement of the proposed procedure and shall be discussed elsewhere.}

In our brane setup, we consider a system $\mathscr S$ with local on-shell stress tensor $T^{ij}(\vect x)$. The set of microstates of $\mathscr S$, at point $\vect x$, with total stress tensor $T^{ij}(\vect x)$ is written as $\{m\}(\vect x)$. We denote the probability for $\mathscr S$ to occupy the microstate $m$ at point $\vect x$ by $p_m(T^{ij}(\vect x);\vect x)$, for which we shall use the shorthand notation $p_m(\vect x)$. Note that, unlike in the Galilean case, the probability depends on the evolution parameter $\vect x$ also through the label $T^{ij}(\vect x)$ of the microstate $m$. The Gibbs entropy is still defined as
\begin{equation}
	S(\{p_m(\vect x)\}) \coloneqq -k_{\textrm B}\sum_m p_m(\vect x)\ln p_m(\vect x),
\end{equation}
where we keep the $\vect x$-dependence explicit everywhere to stress this important matter of fact. The objective function to be maximised in order to find the equilibrium distribution $p_m^*(\vect x)$ is given by
\begin{equation}
	\frac{1}{k_{\textrm B}}G(\{p_m(\vect x)\},\lambda(\vect x)) \coloneqq  -\sum_m p_m(\vect x)\ln p_m(\vect x) - \lambda(\vect x)\left(\sum_m p_m(\vect x)-1\right).
\end{equation}
The Lagrange multiplier $\lambda(\vect x)$ that introduces the pointwise completeness relation for probability density becomes a field on space. Up to this mere change, the extremisation proceeds as before and the equilibrium distribution is reached if all probability densities $p_m^*$ are pointwise equal. If $W(T^{ij}(\vect x))$ quantifies the amount of microstates at point $\vect x$ with total stress tensor $T^{ij}(\vect x)$, the equilibrium distribution obeys
\begin{equation}
	p_m^*(\vect x) = W(T^{ij}(\vect x))^{-1}\qquad\Rightarrow\qquad S(\{p_m^*(\vect x)\}) = k_{\textrm B}\ln W(T^{ij}(\vect x)).
\end{equation}
To establish a notion of temperature in this setting, we consider again two coupled systems and make one of them act as a thermostat by largely increasing its amount of accessible microstates.

Let $\mathscr S_1$ and $\mathscr S_2$ be two systems such that their union $\mathscr S = \mathscr S_1\cup\mathscr S_2$ can be considered as isolated. The interface separating $\mathscr S_1$ and $\mathscr S_2$ is now a space-filling $d$-dimensional hypersurface, that one can consider as constant-$t$ surface up to redefining the $t$-coordinate. For simplicity, we consider that this interface is unique and simply connected. Moreover, we consider that the amount of microstates of $\mathscr S_2$ is infinitely larger than $\mathscr S_1$, hence we shall assume that the following approximation
\begin{equation}
	T_1^{ij}(\vect x)\ll T_2^{ij}(\vect x), \qquad T^{ij}_2(\vect x) = T^{ij}(\vect x) - T^{ij}_1(\vect x) \simeq T^{ij}(\vect x),\qquad \forall\vect x,\ \forall i,j=1,\dots,d \label{eq:constT}
\end{equation}
holds. We again interpret $\mathscr S_1$ as the system of interest and $\mathscr S_2$ as the ``thermostat'' to which it is coupled. Under these hypotheses, the Galilean analysis can be effortlessly translated, \textit{mutatis mutandis}. Similarly to Eq. \eqref{eq:d6}, the total entropy, seen as a function of the stress tensor of $\mathscr S_1$, is given by
\begin{equation}
	 S\!\left(T^{ij}_1(\vect x)\right) = S_1\!\left(T^{ij}_1(\vect x)\right)+ S_2\!\left(T^{ij}(\vect x)-T^{ij}_1(\vect x)\right).
\end{equation}
The equilibrium state, with stress tensor $T^{*ij}_1(\vect x)$, is the most probable state to be occupied by $\mathscr S_1$ and characterised by
\begin{equation}
	\left.\frac{\partial S}{\partial T^{ij}_1(\vect x)}\right|_{T^{*ij}_1(\vect x)} = 0\qquad\Rightarrow\qquad\left.\frac{\partial S_1}{\partial T_1^{ij}(\vect x)}\right|_{T^{*ij}_1(\vect x)} = \left.\frac{\partial S_2}{\partial T^{ij}_2(\vect x)}\right|_{T^{ij}(\vect x) - T^{*ij}_1(\vect x)}. \label{eq:principe zero}
\end{equation}
Equilibrium is therefore characterised by the equivalence of a local and tensorial quantity,
\begin{equation}
	\alpha_{ij}(\vect x)\coloneqq \frac{1}{k_{\textrm B}} \frac{\partial S}{\partial T^{ij}(\vect x)},
\end{equation}
which takes over the role of inverse temperature and is related to the local spaceture as explained in the main text below Eq. \eqref{eq:normalisation CMB}. Equilibrium condition \eqref{eq:principe zero} can thus be elevated as the zeroth principle of Carroll thermodynamics: two systems are at thermodynamic equilibrium if their spacetures equate at each point $\vect x$ of space. Finally, at equilibrium, the probability distribution reads
\begin{equation}
\begin{split}
	p_m\!\left(T^{m,ij}_1(\vect x)\right) = \frac{W_2\!\left(T^{ij}(\vect x)-T^{m,ij}_1(\vect x)\right)}{W\!\left(T^{ij}(\vect x)\right)} &\simeq \frac{1}{Z_c\!\left(\bm\upalpha_2(\vect x)\right)}\exp \left(-\frac{1}{k_{\textrm B}} T^{m,ij}_1(\vect x)\left.\frac{\partial S_2}{\partial T_2^{ij}(\vect x)}\right|_{T^{ij}(\vect x)}\right) \\
	&= \frac{1}{Z_c\!\left(\bm\upalpha_2(\vect x)\right)}\exp \left(-T^{m,ij}_1(\vect x)\alpha_{2,ij}(\vect x)\right).
\end{split} \label{eq:ppp}
\end{equation}
which, by virtue of the equilibrium condition \eqref{eq:principe zero} reveals nothing but the discrete version of the Carroll--Maxwell--Boltzmann distribution. The non-vanishing scalar field
\begin{equation}
	Z_c\!\left(\bm\upalpha(\vect x)\right) = \sum_m \exp \left(-\alpha_{ij}(\vect x) T^{ij}_m(\vect x)\right). \label{eq:zccarr}
\end{equation}
is the canonical partition function at fixed spaceture. Prolonging these derivations to continuous spectra straightforwardly transmutes Eqs. \eqref{eq:ppp}--\eqref{eq:zccarr} into Eqs. \eqref{eq:maxwellboltz} and \eqref{eq:normalisation CMB}.

\providecommand{\href}[2]{#2}\begingroup\raggedright\endgroup

\end{document}